Benchmarking M6 Competitors: An Analysis of Financial Metrics and Discussion of Incentives

The M6 Competition assessed the performance of competitors using a ranked probability score and an information ratio (IR). While these metrics do well at picking the winners in the competition, crucial questions remain for investors with longer-term incentives.  To address these questions, we compare the competitors' performance to a number of conventional (long-only) and alternative indices using standard industry metrics.  We apply factor models to measure the competitors' value-adds above industry-standard benchmarks and find that competitors with more extreme performance are less dependent on the benchmarks. We also uncover that most competitors could not generate significant out-performance compared to randomly selected long-only and long-short portfolios but did generate out-performance compared to short-only portfolios.  We further introduce two new strategies by picking the competitors with the best (Superstars) and worst (Superlosers) recent performance and show that it is challenging to identify skill amongst investment managers. We also discuss the incentives of winning the competition compared to professional investors, where investors wish to maximize fees over an extended period of time.

Matthew J. Schneider (mjs624@drexel.edu), Rufus Rankin (ProShares), Prabir Burman (UC Davis), Alexander Aue (UC Davis)

Section 1:  Introduction

*1.1. Benchmark Indices and Net Asset Values (NAVs)*

By the end of the M6 Competition on February 3$^{rd}$, 2023, 58% of competitors had realized negative returns.  The equal-weighted benchmark was up 0.5% and consisted of 50 stocks and 50 Exchange-Traded Funds (ETFs). Of the 163 teams that completed all 12 months of the M6 Competition, only 28% had a higher realized return than the equal-weighted benchmark, demonstrating the difficulty of the competition and reflecting the challenges all active managers face in beating broad benchmarks.  On the other hand, the competition does not fully capture the scope of running an active investment management offering in the following sense: professional investors ostensibly wish to maximize fees over a long period of time, while a short-term competition may encourage riskier behavior, as the upside of winning the competition dwarfs the potential downside of not winning, and may induce the competitors to engage in less than entirely rational behavior (Bali et al. 2011).  To be clear, this is not a suggestion that the competition is flawed in any way or that all competitors performed extremely, simply an attempt to recognize that the incentives of the competition are somewhat different from those of running a hedge fund and, as such, may encourage different, possibly riskier behavior (Witkowski et al. 2023; Lichtendahl & Winkler, 2007).

Our contributions in the paper are three-fold.  First, we empirically document the performance of M6 competitors using industry-standard financial metrics. Compared to Makridakis et al. (2023), we introduce new metrics that competitors were not rewarded for, in an effort to better understand their performance from a practice perspective.  Second, we apply factor models to competitors' returns and find that competitors who performed in the extremes tended to be less dependent on the benchmark indices.  This implies that a strategy of simply replicating the S&P 500 (or other benchmarks) would not likely result in outstanding performance in the competition. Third, we test the momentum of competitors using two new strategies (Superstars and Superlosers) and compare the competitors' performance to randomly selected portfolios.  The results indicate that neither long-only nor short-only portfolios could produce extreme Sharpe Ratios (or IRs) to win the competition, but the long-short portfolios could. Furthermore, the long-short portfolios were better at controlling downside risks and were mostly uncorrelated with each other.

Figure 1 shows a histogram of the Net Asset Values (NAVs) among competitors, representing the dollar value at the close of the competition on February 3$^{rd}$, 2023. It assumes each competitor started with $100 at the start of the competition.  Notably, only four teams ended with a NAV less than $80 (at least a 20% loss) and only two teams ended with a NAV greater than $120 (at least a 20% gain). Furthermore, 68 (42%) competitors increased their investment during the competition with a NAV exceeding $100, 95 (58%) decreased their investment, and 15 (9%) competitors had identical daily returns and NAVs to the benchmark as illustrated by the large spike in the middle.

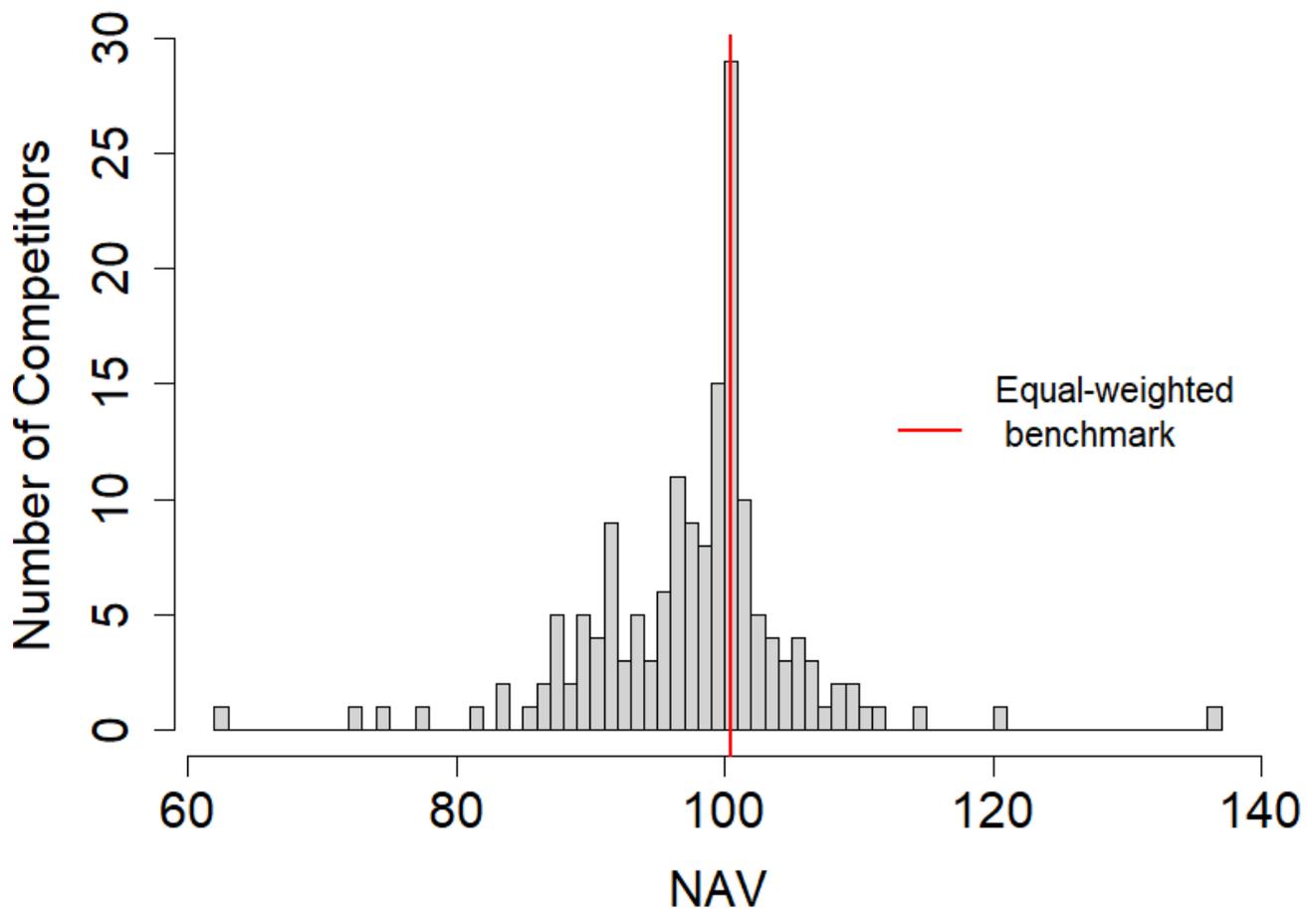

Figure 1: Histogram of the ending Net Asset Values (NAVs) for all competitors when starting with $100

A reasonable starting point for considering relative performance is the S&P 500 and Nasdaq-100, two large-cap indexes representing large-cap U.S. stocks, which many passive investors track via low-cost exchange-traded funds (ETFs).  The majority of assets in the universe for the competition are individual large-cap equities or ETFs that track some set of large-cap equities.  The performance of the competitors relative to these passive indices was encouraging, as 2022 was a very challenging year for equities, and both indices experienced significant drawdowns during the timeframe of the competition.  This suggests that while most competitors did not add value over the equal-weighted portfolio of universe constituents, a majority did manage to express smaller losses than widely used long-only benchmarks. This underlines one of the arguments made for active investment management and the M6 competition: active management can reduce risk, independently of attempting to generate market-beating returns. Another point to note is that, as discussed in DeMiguel (2009), it is extremely difficult to beat a simple equal-weighted allocation with more complex methods.

As the competition was related to active management and allowed competitors to use both long and short positions, it also makes sense to look at active benchmarks. When considering the NAVs of both large-cap and active benchmark indices[1] during this same time period, the S&P500 ended with a NAV of $95.56, NASDAQ-100 with $90.86, Crypto with $62.54, Commodity Trading Advisors (CTA) with $104.26, Equity Long-Short (LS) with $99.99, Event-Driven with $101.33, and Market Neutral with $101.93. A more optimistic look emerges when competitors are compared to the ending NAV of the S&P 500, showing that only 50 (31%) competitors had a lower ending NAV. On average, the M6 competitors ended with a NAV of $97.60. However, besides the ending NAV, the pathways to the ending NAV are usually a major focal point for investors. The NAV can be measured at the close of each trading day as,

$$NAV_t = \$100 \times \prod_{t'=1}^{t'=t}(1+r_{t'})$$

where $t$ is the trading day and $r_{t'}$ is the percent return on trading day $t'$. In this paper, we use raw returns, $r_t$, which differ from log returns used by Makridakis et al. (2023), $r_t^* = \log(1+r_t)$, but are a close approximation. To transform log returns to raw returns, we use the equation $r_t = \exp(r_t^*) - 1$.

Figure 2 plots the daily NAVs of the M6 competition for all competitors, benchmark indices, and three quantiles (5%, 50%, 95%) of competitors. The quantile NAVs are calculated as the $q$% quantile of all 163 competitor NAVs on day $t$. They do not track a single competitor across time. The figure shows that the Crypto index and 16 (10%) competitors are outside the 5% and 95% quantiles, demonstrating a more extreme return stream. Most of the other benchmark indices are within the quantiles, except for the Nasdaq-100, which exceeded the 95% quantile early in the competition and the 5% quantile several times later in the competition. A savvy investor may also question whether the top 5% (or bottom 5%) of competitors exhibit mean reversion behavior – *i.e.*, does a large gain in the past month lead to a large loss in the next month? We investigate this phenomenon in Section 3.2 based on the well-known study (Burghardt et al., 2007), which shows that investing in past superstars may not be optimal.

---

[1] Data was made available to us from https://nilssonhedge.com/index/

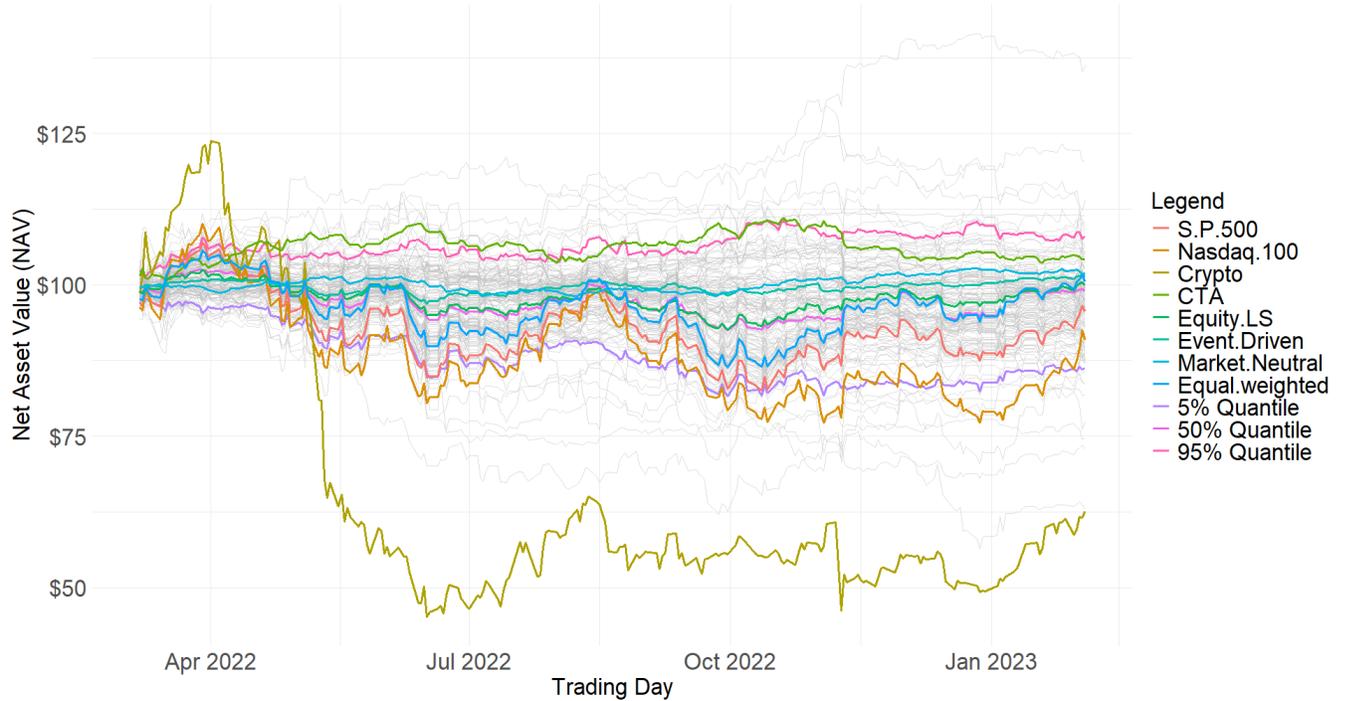

Figure 2: Net asset values (NAVs) of all competitors, benchmark indices, and quantiles for each day of the M6 Competition

*1.2. Financial Metrics*

Investors consider a variety of financial metrics when making investment decisions. The most prominent metric is the annualized Sharpe Ratio (SR), while the M6 Competition (Makridakis et al., 2023) uses the related Information Ratio (IR),

$$IR = \frac{\sum_{t=1}^{t=n} \log(1 + r_t)}{\sqrt{\frac{1}{n-1} \sum_{t=1}^{t=n} (\log(1 + r_t) - \frac{1}{n} \sum_{t=1}^{t=n} \log(1 + r_t))^2}}.$$

where $r_t$ is a portfolio's raw percent return at the close of day $t$ and $n$ is the total number of trading days (238).

The annualized SR is defined as

$$SR = \frac{\prod_{t=1}^{t=n}(1 + r_t)^{n^*/n} - 1}{\sqrt{n^*}\sigma}$$

where $t$ is the trading date, $n^*$ is the total number of periods in a year (252), and $\sigma$ is the standard deviation of daily returns. The numerator of SR is the annualized return (without the log approximation), and the denominator is the annualized standard deviation. As a result, the IR can be over an order of magnitude higher than the SR, but the correlation should be very close to 1 with reasonable returns (if the raw returns are unreasonably negative or positive, e.g., 30% or -30%, then the log approximation can break down resulting in returns of 26.2% and -35.7%, respectively).

The primary difference between the M6 Competition's use of the IR and our use of the SR is that we annualize the expected raw returns and standard deviations to compute SR, while the M6 Competition uses log returns and does not annualize the standard deviation, resulting in larger IR values (the correlation between SR and IR across competitors is 0.996 with only slight differences due to the log approximation). Sharpe (1994) suggests annualizing and compounding multi-period returns when calculating the SR. Regardless of IR or SR, investors still expect a positive value but desire SRs to be closer to 2. SRs exceeding 2 are more theoretically ideal but problematic because they may indicate fraud or trading strategies that miraculously fail at one point in the future. For example, one of Bernie Madoff's feeder funds, Fairfield Sentry Ltd hedge fund, had an annualized SR of 2.47 from December 1990 to October 2008 (Bernard & Boyle, 2009). Also, portfolios that short put options on the S&P 500 can have an unusually high SR until a market downturn. Many portfolio managers deem these strategies as "too good to be true" and avoid them due to perceived risks.

Investors also like to look at a statistic called the maximum drawdown (MDD), which is the maximum percent loss over the entire history of the portfolio, from peak to trough. For example, if Portfolio A started with $100 and had the following consecutive daily NAVs: $110, $200, $100, $150; the MDD would be 50% because the maximum percent loss occurred from $200 to $100, representing a 50% decrease. However, if Portfolio B started with $100 and had the following consecutive daily NAVs: $95, $90, $85, $80; the MDD would be 20%. An investor may prefer Portfolio B due to the lower MDD, even though Portfolio A had a higher return than Portfolio B. The MDD is defined as

$$MDD = \max_{t_1 < t_2} \frac{-(NAV_{t_2} - NAV_{t_1})}{NAV_{t_1}}$$

where $t_1 = 1, \ldots, n$ and $t_2 = 2, \ldots, n$ for all $t_1 < t_2$. Since MDD is highly dependent on time, it greatly depends on when an investor enters a portfolio. Thus, two different investors can have a completely different MDD for the exact same portfolio. Although MDD is a maximum-order statistic lacking in statistical theory (see Section 2.3 for the theoretical explanation), it greatly influences outcomes and investor decision-making. For example, investors may issue a stop loss at an MDD threshold to reduce their overall risk.

Investors can replace the denominator of the SR with MDD to create the Calmar Ratio (CR), but this metric also suffers from similar maximum order statistic problems. In the case of Portfolio A, which has a higher standard deviation of returns and MDD than Portfolio B, decisions may be similar. The signs of the SR and CR are entirely dependent on the expected returns, as they share the same numerator. The CR is defined as

$$CR = \frac{Annualized\ Return}{MDD}$$

where the annualized return is the same as the numerator in the SR.

Investors are also concerned about the duration of any drawdown until recovery to the prior peak, producing more interesting metrics, such as the Ulcer Index (UI) (Martin & McCann, 1989) and the Ulcer Performance Index (UPI). UI measures the pain (duration and depth) of investor losses, from peak to trough to peak of a portfolio's drawdowns,

$$UI = \sqrt{\frac{1}{n}\sum_{t=1}^{t=n} DD_t^2}$$

where $DD_t$ is the drawdown on trading day t since the previous peak. UI is a practical metric because investors are more likely to withdraw their assets the longer a portfolio is underwater. For example, the UI of Portfolio C with NAVs of $100, $90, $100, $110 would be 0.058 (with drawdowns of 10%, 0%, and 0%), while the UI of Portfolio D with NAVs of $100, $90, $90, $110 would be 0.082 (with drawdowns of 10%, 10%, and 0%), representing a greater ulcer for an investor in Portfolio D. Lastly, the UPI replaces the denominator of the SR with the UI,

$$UPI = \frac{Annualized\ Return}{UI}.$$

Not all volatility is bad. For example, two portfolios may have identical negative returns but different positive returns. In this case, the portfolio with the larger positive returns has a higher standard deviation, which any investor prefers. To account for this, we measure each portfolio's upside volatility ($vol^+$) and its downside volatility ($vol^-$). Upside volatility is defined as

$$vol^+ = \sqrt{\frac{\sum_{t=1}^{t=n_1}\left(r_t^+ - \frac{\sum_{t=1}^{t=n_1} r_t^+}{n_1}\right)^2}{n_1}}$$

where $r_t^+$ denotes the positive raw returns and $n_1$ is the sample size of the positive returns. Similarly, downside volatility is defined as

$$vol^- = \sqrt{\frac{\sum_{t=1}^{t=n_2}\left(r_t^- - \frac{\sum_{t=1}^{t=n_2} r_t^-}{n_2}\right)^2}{n_2}}$$

where $r_t^-$ denote the negative raw returns and $n_2$ is the sample size of the negative returns.

Note that the case of $vol^+ \gg vol^-$ is not accounted for in the denominator of the SR, but it is more likely to occur when the numerator of SR is positive. Investors prefer $vol^+ \gg vol^-$ and in such a case, the MDD is likely to be smaller.

Table 1 presents the financial metrics for the major benchmark indices, including the equal-weighted benchmark in the M6 Competition.[2] Results are presented in increasing order of IR and SR (with a risk-free rate of 0) over the $n = 238$ trading days of the competition. SRs were relatively moderate (between -1 and 1), and SRs were positive for those indices that gained money ($NAV_{238} > \$100$). The relationship between the SR and CR was also monotonic, indicating that the MDDs were monotonically increasing with the annualized standard deviation of returns. MDD and UI were also monotonically increasing with each other, indicating that the lengths of the drawdowns were not greatly mismatched to the depths of the drawdowns.

---

[2] For the analysis in this paper, we removed the Crypto returns that occurred on the weekends (non-trading days) and assumed that all benchmark indices and competitors traded on the same 238 trading days for a fair comparison. Crypto had an average daily return of -0.115% when including all 334 days that spanned the competition, and an average daily return of -0.126% when only including the 238 trading days.

|  | IR | SR | MDD | Autocorrelation | vol$^+$ | vol$^-$ | CR | UI | UPI |
|---|---|---|---|---|---|---|---|---|---|
| Crypto | -12.32 | -0.66 | 0.64 | -0.04 | 0.02 | 0.03 | -0.62 | 0.63 | -0.62 |
| Nasdaq-100 | -4.77 | -0.30 | 0.30 | -0.05 | 0.01 | 0.01 | -0.32 | 0.21 | -0.46 |
| S&P 500 | -3.02 | -0.20 | 0.23 | -0.01 | 0.01 | 0.01 | -0.21 | 0.14 | -0.33 |
| Equity LS | -0.01 | 0.00 | 0.10 | -0.02 | 0.00 | 0.00 | 0.00 | 0.05 | 0.00 |
| Equal-weighted | 0.45 | 0.03 | 0.18 | 0.04 | 0.01 | 0.01 | 0.03 | 0.10 | 0.06 |
| Event Driven | 5.40 | 0.36 | 0.04 | 0.05 | 0.00 | 0.00 | 0.35 | 0.02 | 0.92 |
| Market Neutral | 9.18 | 0.62 | 0.03 | 0.17 | 0.00 | 0.00 | 0.64 | 0.02 | 1.34 |
| CTA | 9.32 | 0.64 | 0.07 | 0.18 | 0.00 | 0.00 | 0.69 | 0.04 | 1.28 |

Table 1: Financial Metrics of the Major Benchmark Indices

Table 2 shows that most competitors had a negative SR (with NAVs below $100), and the SRs were between -1.76 and 2.58. Furthermore, some competitors had high UI values, illustrated by some of the gray NAV lines in the bottom part of Figure 2. The values of first-order autocorrelation seemed to be moderate, but some competitors had high MDDs.

|  | IR | SR | MDD | Autocorrelation | vol$^+$ | vol$^-$ | CR | UI | UPI |
|---|---|---|---|---|---|---|---|---|---|
| Min. | -29.18 | -1.76 | 0.01 | -0.13 | 0.00 | 0.00 | -1.00 | 0.01 | -2.44 |
| 1st Qu. | -8.37 | -0.53 | 0.10 | 0.00 | 0.00 | 0.00 | -0.47 | 0.05 | -0.81 |
| Median | -1.34 | -0.09 | 0.16 | 0.04 | 0.01 | 0.01 | -0.11 | 0.08 | -0.22 |
| Mean | -3.06 | -0.18 | 0.15 | 0.02 | 0.01 | 0.01 | -0.04 | 0.08 | 0.04 |
| 3rd Qu. | 1.02 | 0.07 | 0.18 | 0.05 | 0.01 | 0.01 | 0.07 | 0.10 | 0.13 |
| Max. | 32.89 | 2.58 | 0.47 | 0.16 | 0.02 | 0.02 | 4.95 | 0.36 | 11.91 |

Table 2: Financial Metrics of the M6 Competitors

1.3. Management and Performance Fees

Another primary consideration for investors is the effect of management and performance fees on portfolio performance. Management fees are an annualized flat percentage fee collected off the annual NAV. Hedge funds collect these monthly or quarterly, typically between 1% and 2% annually. Performance fees are incentive fees collected on the increase in NAV from an initial NAV, typically between 10% and 20%. Elton et al. (2003) found that funds with incentive (performance) fees have higher average risk-adjusted performance than funds without incentive fees, which suggests that they tend to attract superior managers. In the context of the design of the M6 Competition, the competitors did not charge management or performance fees, but we analyzed the effect of moderate (1%/10%) fees.

Management and performance fees are difficult to calculate since the performance fee is only incurred on new high-water marks (HWMs). For example, if Portfolio A went from a $100 NAV to a $200 NAV (new HWM), there would be a 10% performance fee on the gain (+$100) for a total of $10, reducing the new NAV from $200 to $190. However, if portfolio A had a subsequent drawdown, the performance fee would not be incurred until a new NAV exceeded the previous HWM of $200. For this reason, the performance fee calculation is highly dependent on when an investor invests and is usually outsourced to an accounting firm. For simplicity, with the short timeline of the M6 Competition, we assume a low fee structure (1%/10%) and deduct performance fees and prorated management fees daily.

Figure 3 plots the 5%, 50%, and 95% quantiles of the NAVs for all 163 competitors in the M6 competition. NAVs are shown before fees (blue, dark blue, and black for the 5%, 50%, and 95% quantiles, respectively) and after fees (green) by assuming 1% management and 10% performance fees. The difference between non-fee and fee portfolios heightens as the competition length increases. Since the competition lasted for only 238 trading days, the net effect of the 1%/10% fees was not substantial. For the median competitor, the 1%/10% fees decreased the SR from -0.09 to -0.17, the IR decreased from -1.3 to -2.6, and the MDD increased from 15.5% to 16.0%. However, if the competition lasted several years, the effects of management and performance fees would be much more severe. Unlike the short-term nature of the competition, the portfolio managers' goal is to retain assets under management (AUM) and maximize these fees over a long period of time.

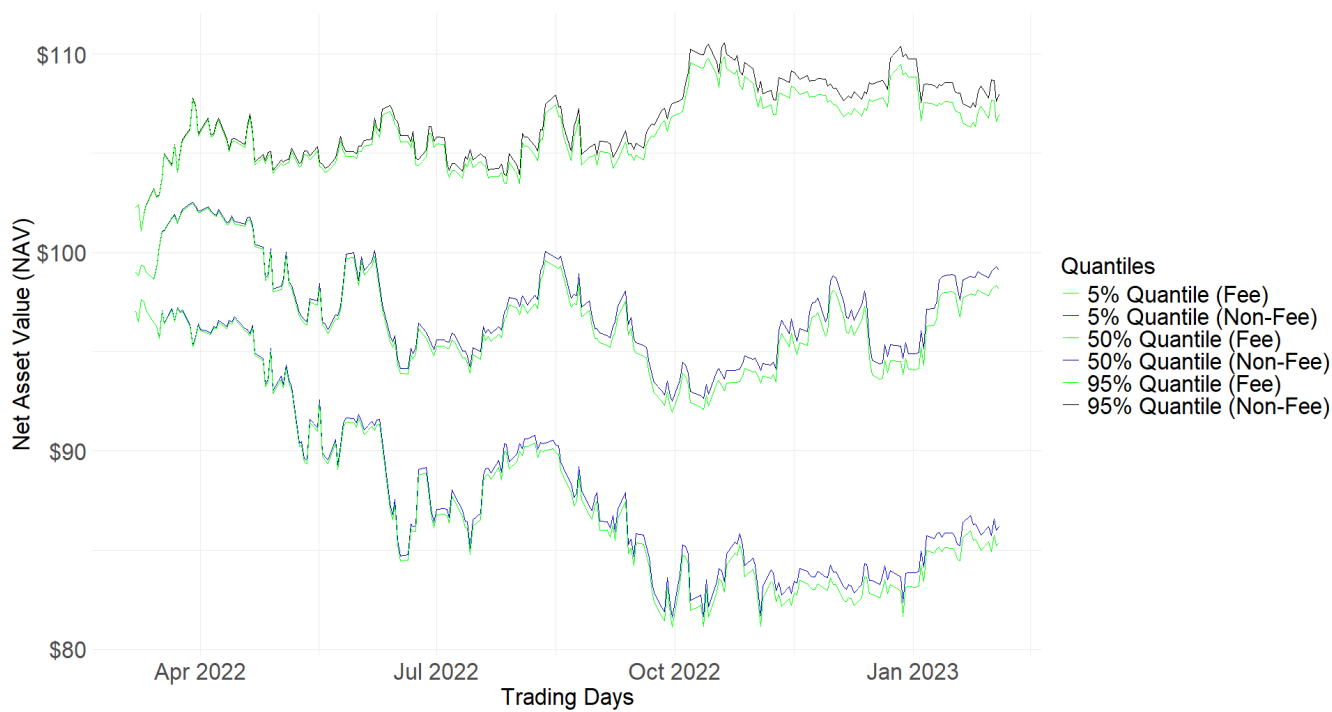

Figure 3: The Effect of Management and Performance Fees on Competitor NAVs

Section 2. Statistical Analysis of the M6 Competitors

*2.1 Statistical Analysis of a Competitor's Value-Add*

The statistical performance is assessed by measuring the value-add (alpha) of competitors' portfolios compared to major indices. Alpha is frequently mentioned on a portfolio's marketing sheet and represents the independent return that a portfolio produces above and beyond benchmark indices.

From a statistical perspective, alpha is the estimated intercept of a multiple linear regression (or factor model). It represents the excess return over the explanatory power of the predictor (X) variables. The predictor variables are typically benchmark indices or other factors found in the suite of Fama-French regression models (Fama & French, 1995). Intuitively, alpha represents the additional return of a portfolio when all benchmark indices produce a 0 return. Investors may compare benchmark indices with minimal fees, such as the S&P 500 or the Nasdaq-100, to the focal portfolio. The idea is that investors would not want to pay management and performance fees if a portfolio cannot produce excess returns independent of the minimal fee benchmark indices. The standard approach to identify which portfolios have positive alphas is "to perform many individual statistical tests on the alphas of the factors or funds relative to the benchmark model, and then make a selection based on the significance of these individual tests" (p. 3457, Giglio et al. 2020).

For the factor model, let $r_t$ be the daily return of a competitor's portfolio on day $t$, $r^f{}_t$ be the de-annualized risk-free rate at day $t$, $\alpha$ be alpha, $f_{j,t}$ be the daily return of a benchmark index (factor) on day $t$, $\beta_j$ be its coefficient, and $\varepsilon_t$ be the standard error term that is i.i.d. with mean 0 and constant variance from a normal distribution,

$$r_t - r^f{}_t = \alpha + \sum_{j=j}^{j=J} \beta_j(f_{j,t} - r^f{}_t) + \varepsilon_t.$$

For a portfolio manager to market their portfolio effectively, alpha should be positive. This desired criterion creates a selection bias for investors because portfolio managers rarely advertise portfolios with negative alphas. Furthermore, portfolio managers may introduce additional biases to present a more favorable alpha, such as changing the benchmark indices, altering the time windows, reducing management and performance fees, or ignoring the lack of statistical significance. A major advantage to the M6 Competition structure is that most of these biases can be controlled using the same benchmark indices, fees, time windows, and statistical significance calculations. In addition, none of the data is missing which tends to pose problems for the computation of alphas (Giglio et al. 2020).

The Appraisal Ratio (AR) (Hecht, 2014), which divides a portfolio's alpha by the standard deviation of the factor model's residuals, is lesser known. The AR is a risk-adjusted measure of alpha and is shown below:

$$AR = \frac{\alpha}{sd(\varepsilon_t)}$$

Hecht (2014, p. 5) also cites a mathematical proof that shows that the maximum attainable SR of a total portfolio is a function of the AR of a new asset and the existing SR of the total portfolio. This implies that for portfolio optimization, a higher AR from a new asset translates to a higher SR for the total portfolio. Statistically, although the AR divides annualized alpha by the annualized standard error of the residuals, a proper z-statistic divides the raw alpha by its standard error. We note that the correlation between the ARs and the z-statistics is 1.

We applied the factor model to all 163 competitor portfolios using the seven benchmark indices and one benchmark index (the S&P 500). The motivation behind using the one factor model is that competitors only had access to investable assets in the S&P 500 during the competition, and this represents a fair comparison. However, we also include the seven benchmark indices as factors in a seven factor model to check for robustness to the broader market, which includes strategies like Equity Long-Short (LS) that competitors could have replicated. We note that using seven benchmark indices may be arbitrary, resulting in different alpha, p-values, or AR values. However, we test for robustness between the two models and note that modern investors now have many potential benchmarks accessible via low-cost mutual funds or ETFs, and the hurdles for active managers are higher. Nevertheless, we apply both models to compare the results.

Table 3 presents the results for the seven factor model. Of the 163 alphas in the seven factor model, 12 alphas were statistically significant at the 0.05 level, including the largest (annualized) alpha of 30.8%. However, 11 of the 12 statistically significant alphas were negative. For comparison, the equal-weighted portfolio in the M6 competition had an alpha of 3.4% with a p-value of 0.26 and an AR of 1.16.

|  | Value-add | | | Factor Model Coefficients | | | | | | |
| --- | --- | --- | --- | --- | --- | --- | --- | --- | --- | --- |
|  | Annualized Alpha[3] | p-value | Appraisal Ratio[4] | S&P 500 | NASDAQ-100 | Crypto | CTA | Equity LS | Event Driven | Market Neutral |
| Min. | -0.41 | 0.00 | -3.23 | -2.23 | -1.47 | -0.05 | -0.82 | -2.86 | -2.60 | -2.31 |
| 1st Qu. | -0.07 | 0.21 | -1.07 | 0.05 | -0.30 | -0.01 | -0.14 | -0.22 | -0.20 | -0.31 |
| Median | -0.02 | 0.37 | -0.25 | 0.39 | -0.16 | 0.00 | 0.03 | 0.29 | 0.16 | -0.07 |
| Mean | -0.03 | 0.44 | -0.24 | 0.35 | -0.14 | 0.00 | 0.07 | 0.35 | 0.13 | 0.02 |
| 3rd Qu. | 0.03 | 0.69 | 0.66 | 0.63 | 0.00 | 0.01 | 0.23 | 1.06 | 0.38 | 0.33 |
| Max. | 0.31 | 1.00 | 2.16 | 2.70 | 1.28 | 0.05 | 1.27 | 5.13 | 2.82 | 2.53 |

Table 3: Summary Statistics of Annualized Alphas, Annualized ARs, and Coefficients for the Seven Factor Models

Table 4 presents the results for the one factor model using the S&P 500 only and alphas are similar with the seven factor model. Of the 163 alphas in the one factor model, 5 alphas were statistically significant at the 0.05 level, including the largest (annualized) alpha of 30.0%. However, 4 of the 5 statistically significant alphas were negative.   For comparison, the equal-weighted portfolio in the M6 competition had an alpha of 3.1% with a p-value of 0.47 and an AR of 0.73.

---

[3] The factor model estimates of alpha on a daily level. Alphas are annualized by multiplying the intercept by the number of trading days (238) in the competition, but the industry standard is normally 252 days.

[4] ARs are annualized by dividing the annualized alphas by the annualized standard deviation of returns.

|  | Value-add | | | Factor Model Coefficient |
| --- | --- | --- | --- | --- |
|  | Annualized Alpha | p-value | Appraisal Ratio | S&P 500 |
| Min. | -0.42 | 0.00 | -3.19 | -0.76 |
| 1st Qu. | -0.06 | 0.29 | -0.87 | -0.01 |
| Median | -0.01 | 0.47 | -0.16 | 0.32 |
| Mean | -0.02 | 0.51 | -0.21 | 0.31 |
| 3rd Qu. | 0.03 | 0.75 | 0.49 | 0.68 |
| Max. | 0.30 | 0.99 | 2.06 | 1.32 |

Table 4:  Summary Statistics of Annualized Alphas, Annualized ARs, and Coefficient for the One Factor Models

*2.2 Autocorrelation, R-Square, and Joint Test of Equality*

To check the autocorrelation of the competitors' returns, we apply a Durbin-Watson test and find that 20 of 163 competitors exhibit statistically significant autocorrelation. We use the R package prais (version 1.1.2) and applied the Prais-Winsten estimator to estimate the coefficients of the one factor model in the presence of autocorrelation.  Results are fairly similar and there are still 5 statistically significant alphas, with 4 of 5 negative alphas.  However, the original competitor with the highest alpha value keeps a similar alpha when rounded (30.0%), but its p-value shifts from 0.041 to 0.0501. Additionally, a new competitor with a positive alpha value of 7.4% (rounded) became significant with its p-value shifting from 0.065 to 0.035 after the autocorrelation correction.  Overall, the one factor model is fairly robust to the autocorrelation in the competition dataset.

We also compare the R-square values of the one factor model to the competitors' SR and alpha values. We find that those competitors with extreme alphas (and high values of SR) all had R-square values below 50% (and frequently below 20%). The 95% confidence interval for the correlation between absolute value of alpha and R-square was (-0.48, -0.22), indicating that larger values of R-square were associated with alphas closer to 0.   The same held true when comparing R-square to the SRs, which are not dependent on the output of a factor model.  The 95% confidence interval for the correlation between the absolute value of the SR and R-square was (-0.62, -0.41), indicating that larger value of R-square were associated with SRs closer to 0.   Overall, these results imply that competitors who performed in the extremes tended to be less dependent on the benchmark index.  Similar findings also held true when using the seven factor model and Figure 4 displays the results for the one factor model.

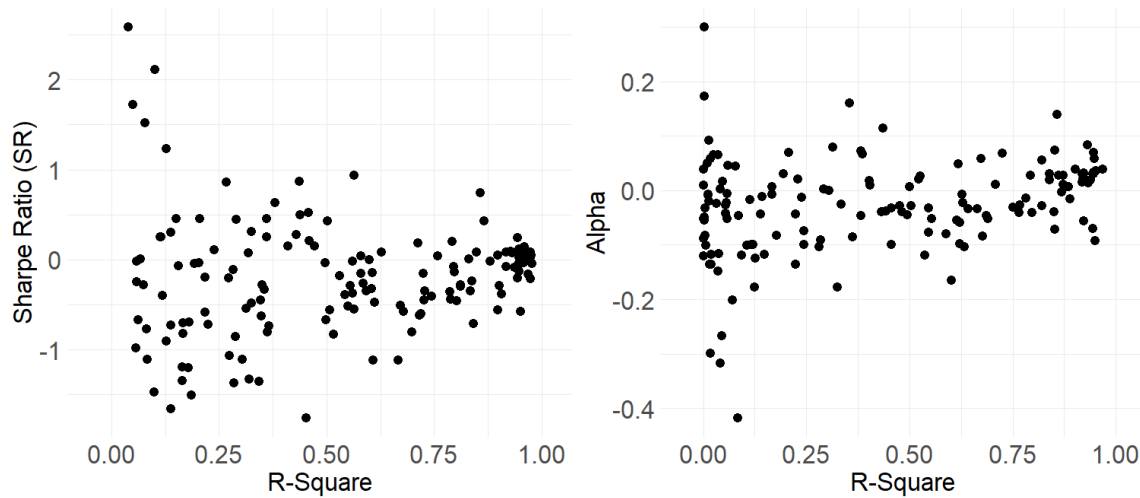

Figure 4: Scatter Plot of R-Square vs. Sharpe Ratio (SR) and Alpha, Using a One Factor (S&P 500) Model

Finally, we perform a Gibbons, Ross, Shanken (GRS) test (Gibbons et al. 1989) for the joint equality of all alphas being equal to 0. The null hypothesis is

$$H_0: \alpha_1 = \alpha_2 = \cdots = \alpha_{163} = 0$$

and it is rejected when there is at least one alpha not equal to 0. We implement the GRS test using the R package GRS.test (version 1.2) for both the one factor and seven factor models. For the one factor model, the test statistic is 15.3 with a p-value < 0.0001, rejecting the null hypothesis that all alphas are equal to 0. For the seven factor model, the test statistic is 15.1 with a p-value < 0.0001, rejecting the null hypothesis that all alphas are equal to 0. Since both tests rejected their null hypotheses, we conclude that at least one of the competitor's alphas is different than 0.

*2.3 Theoretical Commentary on the Maximum Drawdown (MDD)*

As mentioned, investors use the MDD to manage risk and influence decision-making. The MDD calculates the maximum loss an investor experiences over time and is a maximal order statistic. It is the maximum percent loss over any $m \leq n$ consecutive trading days. In our case, we assume that the returns from each of the $n$ trading days in the competition are independent and identically distributed (i.i.d.), which we believe is a realistic assumption. Thus, the percent loss of $m$ consecutive days can be represented as the one minus the product of $m$ consecutive returns, and its maximum is a simplification of the MDD equation when the $m$ days occur from $t_1 + 1$ to $t_2$.

Theoretically, for maximal order statistics, if there are n observations which are i.i.d. with a common cumulative distribution function $F$, then the behavior of any middle order statistic $j$ is rather simple and straightforward. Its mean is about $F^{-1}(j/(n+1))$ with variance of order $1/n$. However, the distribution of the extreme order statistics, say the largest, can be complicated as it depends on the common distribution $F$. For the standard normal distribution, its mean is order square root of $log(n)$ and variance is of order $1/log(n)$. For the exponential distribution, the mean is of order $log(n)$ with finite variance. For distributions with Pareto tails, both the mean and the variance grow algebraically with the sample size $n$. These results point to two aspects of the largest order statistic, such as the MDD. First, the mean grows as the sample size grows. Second, the largest order statistic is unstable as its variance can decrease to zero slowly (normal case), or a non-vanishing constant as in the exponential case, or slowly exploding as in the Pareto case. A good account of these results can be found in the book by David and Nagaraja (2003). For stationary sequences, the behavior of the middle order statistics is stable, as in the i.i.d. case. However, extreme order statistics can exhibit unstable behavior, and a good deal of technical work is available, *e.g.*, Hsing (1988).

In the case of nonstationary time series, ARIMA modeling is quite standard for nonstationary time series with trend, *e.g.*, Box, Jenkins and Reinsel (2008). However, there are some potential problems. If the plot of the data looks like a trend plus stationary noise with non-negligible noise/variance, then modeling the first difference (or the second difference) of the observed series as an invertible ARMA may not be appropriate. If an ARMA series is invertible, then its cumulative sum tends to behave like a smooth process (like a Brownian motion) over time. Thus, modeling the first difference of sequence with trend as an ARMA may lead to incorrect data analysis.

Section 3. Randomness and Combinations of the M6 Competitors

*3.1 Randomly Selected Portfolios vs. M6 Competitors*

In 1973, Burton Malkiel infamously declared, "A blindfolded monkey throwing darts at a newspaper's financial pages could select a portfolio that would do just as well as one carefully selected by experts." (Malkiel, 1973). Since then, a few investment companies have reproduced these results and found that, on average, 98 out of 100 monkey portfolios performed better than capitalization-weighted stock portfolios each year (Ferri, 2012). Warren Buffet reiterated these beliefs in 2022 by stating in his annual meeting that "you can have monkeys throwing darts at the page and, you know, take away the management fees and everything, I'll bet on the monkeys" (as cited in Koebler, 2022).

We apply Malkiel's hypothesis to explore whether randomly selected portfolios can perform as well as the M6 competitors. We restrict the portfolios to the universe of 100 assets available in the M6 Competition. Each randomly selected portfolio is constructed as a simulation that chooses 10 assets (without replacement) at random with equal-weighting (10% each) for the 12 submission points of the competition.[5] 1,000 simulations are performed for long-only, short-only, and long-short portfolios. Long-short portfolios have an equal chance of being long or short on an asset at each of the 12 submission points. Once a random portfolio is established, it trades on all trading days until the next submission point, when assets and long-short positions are rerandomized. Each day, we assume an equal investment is made for each randomly chosen asset, so that we can average the daily returns across assets to produce a daily return. This process continues until the end of the competition on 2023-02-03.

---

[5] The assets in the portfolios were chosen on 2022-03-06, 2022-04-03, 2022-05-01, 2022-05-29, 2022-06-26, 2022-07-24, 2022-08-21, 2022-09-18, 2022-10-16, 2022-11-13, 2022-12-11, and 2023-01-08.

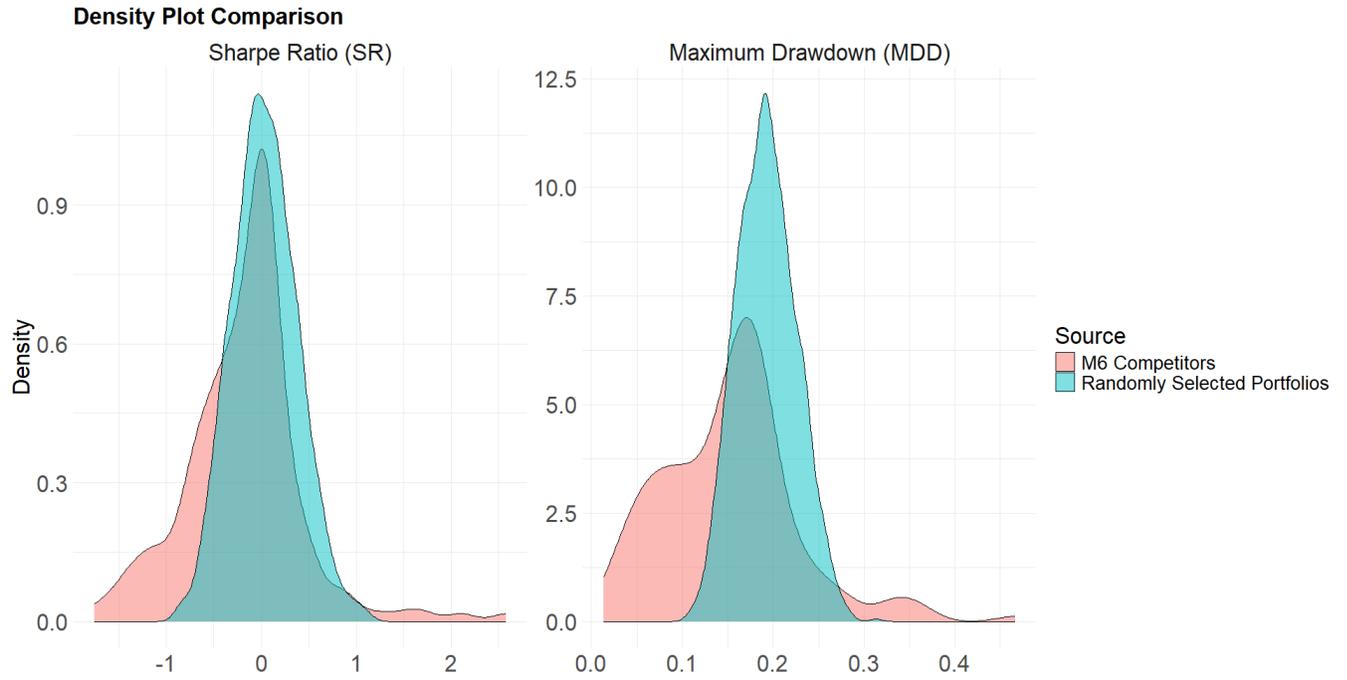

Figure 5: Empirical Density Plot Comparison between M6 Competitors and 1,000 Long-Only Randomly Selected Portfolios

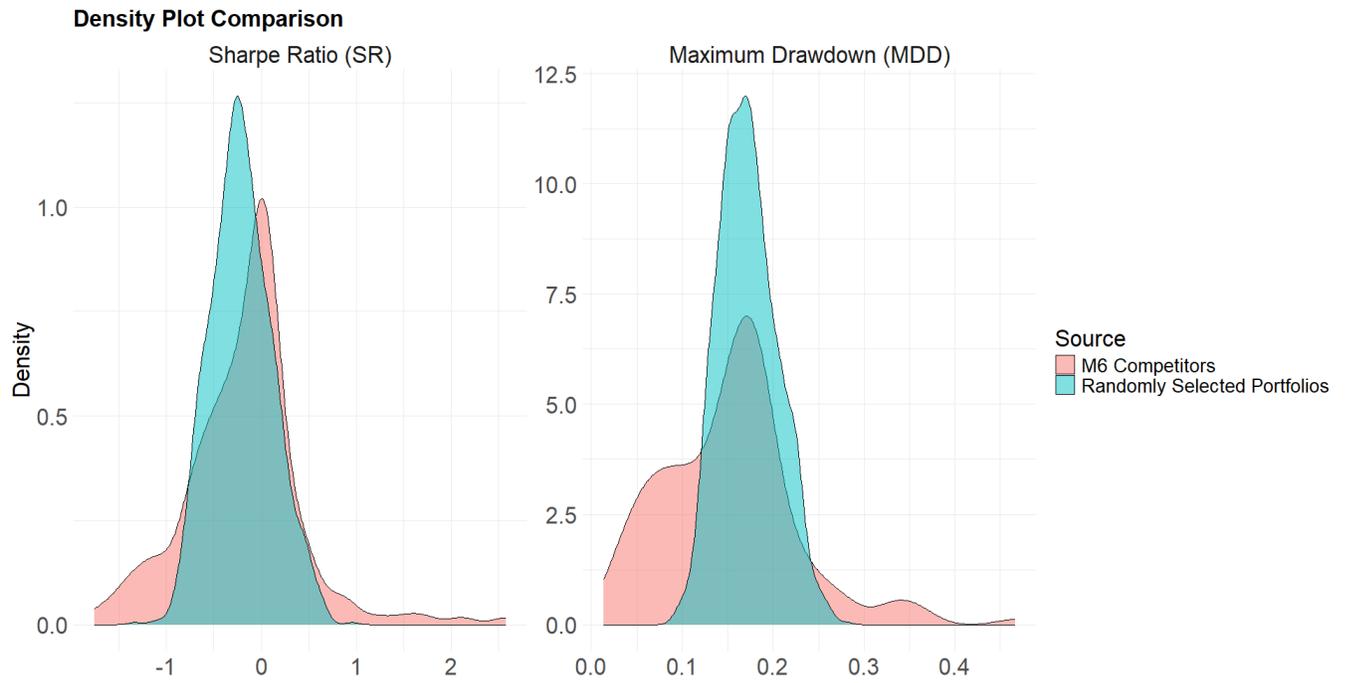

Figure 6: Empirical Density Plot Comparison between M6 Competitors and 1,000 Short-Only Randomly Selected Portfolios

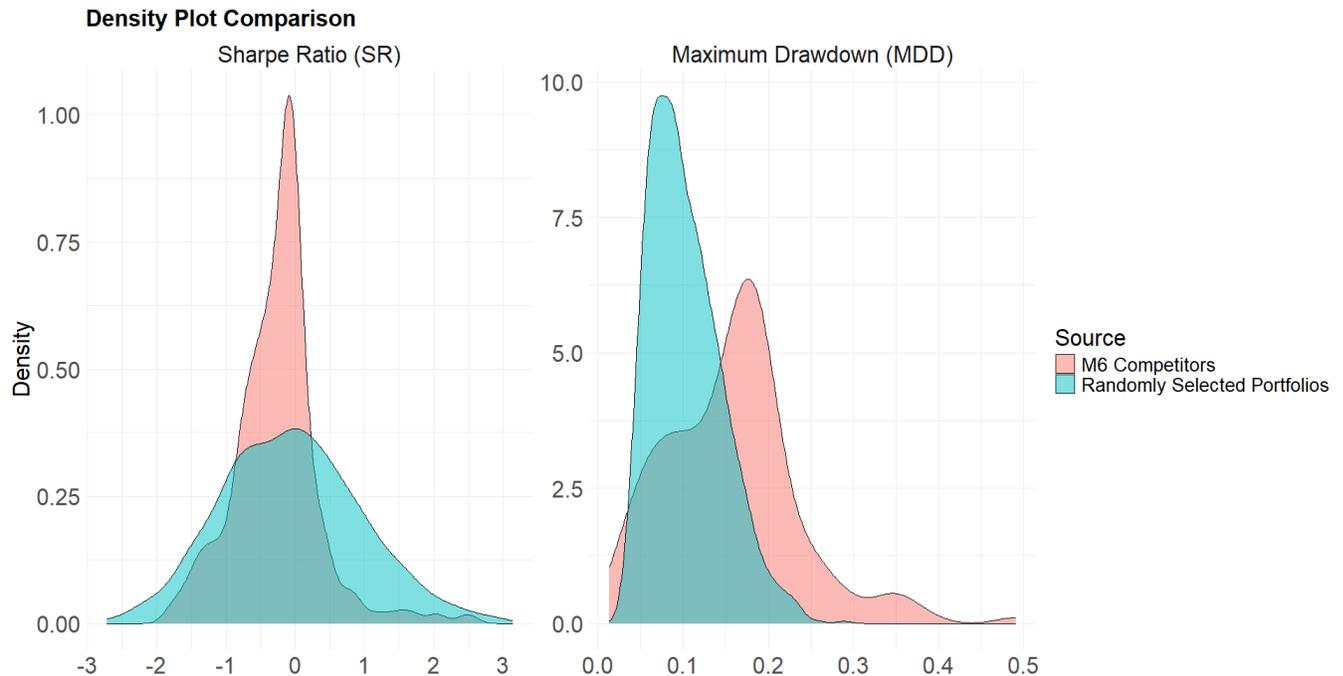

Figure 7: Empirical Density Plot Comparison between M6 Competitors and 1,000 Long-Short Randomly Selected Portfolios

Figures 5, 6, and 7 illustrate the values of SR and MDD for the long-only, short-only, and long-short portfolios, respectively. The long-only portfolios had higher SRs than most competitors but did not exceed the winners of the M6 Competition with SRs above 2. Also, the long-only randomly selected portfolios had higher MDDs and volatility than most competitors, which caused the tighter range of SRs around 0. As a group, the 1,000 long-only portfolios netted a positive return with an average ending NAV of $100.67. This is compared to the M6 competitors, who had an average ending NAV of $97.60, with half of them having an ending NAV above $99.10.

For the short-only portfolios, the widths of the densities were similar to the long-only densities, but the average ending NAV netted a negative return at $96.03. However, on average, the MDDs were lower in the short-only portfolios (17.2% average) than the long-only portfolios (19.3% average).

On the other hand, the long-short randomly selected portfolios in Figure 7 had a wider range of SRs than M6 competitors, providing empirical evidence that a long-short strategy could produce more extreme SRs needed to win a competition. Interestingly, the long-short randomly selected portfolios also had smaller MDDs than M6 competitors, possibly due to an averaging out effect with both long and short positions. The long-short strategy can have MDD advantages during both up and down markets. Compared to the Equity Long-Short (LS) benchmark index with a MDD of 9.6% and ending NAV of $99.99, the long-short portfolios had a comparable average MDD of 10.3% and a median MDD of 9.6%. However, the 1,000 long-short portfolios netted a slightly worse negative return with an average ending NAV of $99.59.

Since the strategy was random, the objective of the randomly selected portfolios was not to win the competition. This may partially explain why the 1,000 long-only, short-only, or long-short portfolios did not have MDDs as extreme as some of the competitors. Theoretically, Nieken and Sliwka (2010) elaborate on risk-taking behavior in competitions with a formal tournament model and find that when the outcomes are uncorrelated, the leading competitors choose a safe strategy more often and the trailing players almost always choose a risky strategy. Also, when the outcomes are correlated and competitors have access to similar strategies, the leading competitor is more likely to choose a riskier strategy to protect their lead. To win a prize in the M6 competition, competitors needed to demonstrate extreme performance on the IR, potentially risking a higher MDD.

To help explain the difference in performance for the randomly selected portfolios, we analyze the correlation structure between their simulated returns. For the long-only portfolios, the first principal component of the 1,000 simulations explained 87.4% of the variation in the simulated returns and the average absolute value of the correlation between long-only strategies was high at 0.87. For the short-only portfolios, the first principal component of the 1,000 simulations explained 87.5% of the variation in the simulated returns and the average absolute value of the correlation between long-only strategies was high at 0.87. For the long-short portfolios, the results were very different. The first principal component of the 1,000 long-short portfolios explained only 7.7% of the variation in the simulated returns and the average absolute value of the correlation between long-only strategies was low at 0.14. Compared to the case of the M6 competition, the first principal component explained 57.7% of the variance in the competitors' returns and the average absolute value of correlations between competitors was moderate at 0.41. These results may help explain why the long-short portfolios had a wider SR distribution than M6 competitors, while using the long-short balancing effect to limit MDD during market cycles.

Although some competitors did not actively invest (e.g., 15 competitors replicated the equal-weighted benchmark), many had higher downside risks than randomly chosen portfolios. Dijk et al. (2014) note that underperformers may take excess risks to increase their rank, while overperformers may take tail risks to protect their rank. They do this by choosing positively skewed assets and negatively skewed assets, respectively. These behaviors fundamentally differ from the fiduciary duties required by law, where non-active investment is not viable, and extreme performance is strongly disincentivized with costly downsides (*e.g.*, losing money, losing investors and AUM, reputational risks, Security and Exchange Commission (SEC) investigations, and investor lawsuits). This incentive to strive for extreme performance is quite different from most asset managers – who often find that they are expected to deliver a small yet positive tracking error relative to their benchmark.

*3.2 Combinations of Competitors: Superstars vs. Superlosers*

Sometimes, a team of high-performing competitors may underperform a carefully selected competitor team. Burghardt et al. (2007) found that a team of uncorrelated portfolio managers outperformed a team of portfolio managers with the highest SRs (Burghardt & Walls, 2011). This decision problem is common in the fund of (hedge) funds (FoFs) space where a FoF's manager decides allocations of investments to several individual portfolios or hedge funds. Interestingly, they also found that "firing team players for poor individual performance and replacing them with managers with higher Sharpe ratios seriously degraded the performance" (pp. 1, Burghardt et al., 2007).

To test this scenario in the M6 Competition, we create two strategies for FoFs seeking to invest in M6 Competitors. The first strategy picks the top 10 performing competitors ('Superstars') from the prior month for an equal-weighted investment during the next month. The second strategy picks the bottom

10 performing ('Superlosers') competitors from the prior month for an equal-weighted portfolio during the next month.

FoFs typically reinvest quarterly because there are fixed switching costs between portfolios. We employ a FoF approach to simulate this exercise, but assume monthly reinvesting at the M6 Competition selection dates to ensure a synchronized and fair comparison with the competition. For realism, we also assume that the competitors' underlying positions were unknown, and we only allow direct investments to competitors. We also do not permit competitors to charge management and performance fees since the effects were minimal in this short competition.

*3.2.1 Superstars Strategy*

Figure 8 shows the NAV of the top 10 superstars during each month of the competition, where we assume that each competitor starts with $100 each month. The figure also plots an average of the top 10 superstars (bold font, 'Superstar Average') with the realized returns from the previous month. The team of Superstars is selected before the selection dates of the competition by choosing the 10 competitors with the highest returns from the previous month (gray lines in the figure). This process starts in month two and is repeated for every subsequent month. The blue line represents the Superstar strategy and shows a steady decline in the NAV. We find an ending NAV of $87.46 in the Superstars strategy, indicating a mean reversion behavior of the prior month's Superstars.

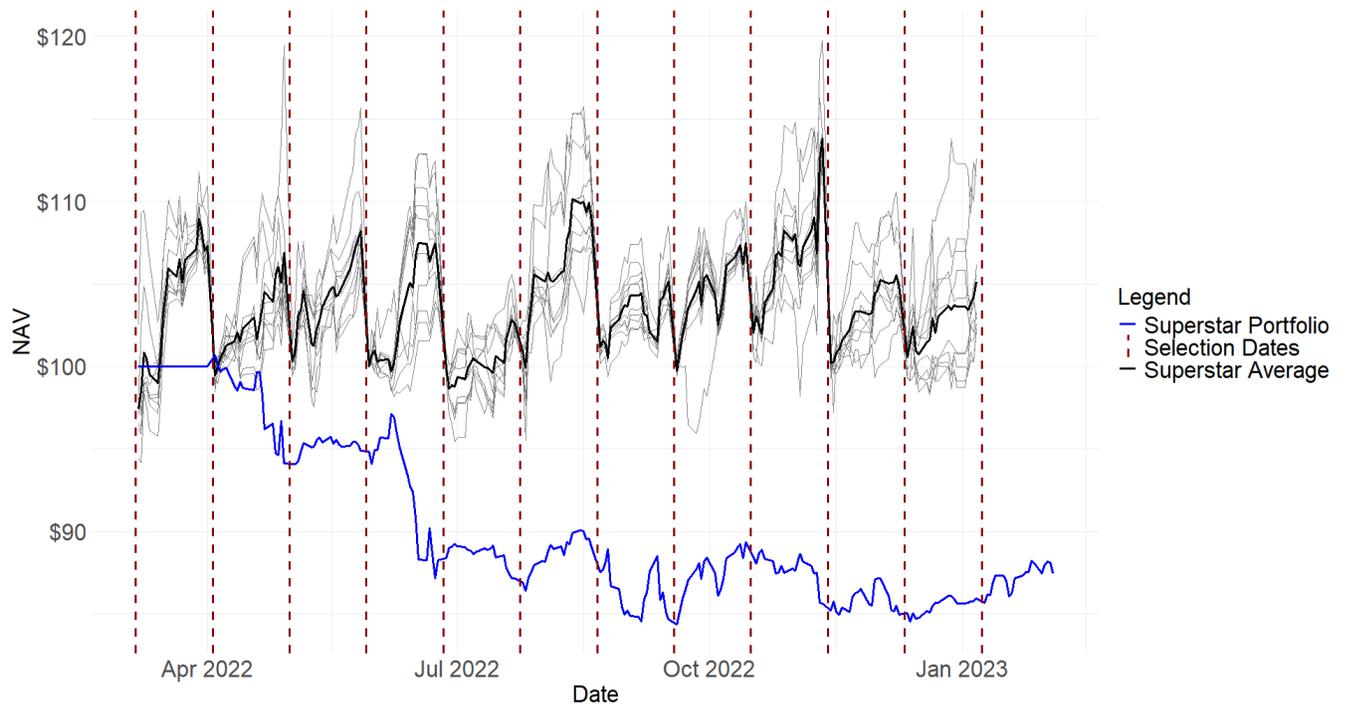

Figure 8: Superstars strategy: Investing in the top 10 winners of the prior month

*3.2.2 Superlosers Strategy*

The process is repeated for the Superlosers strategy, except that the Superlosers consist of the bottom 10 performing competitors of the previous month. Figure 9 shows an ending NAV of $106.95 using the Superloser strategy, indicating that the prior month's top losers tend to increase their NAVs in the next month.

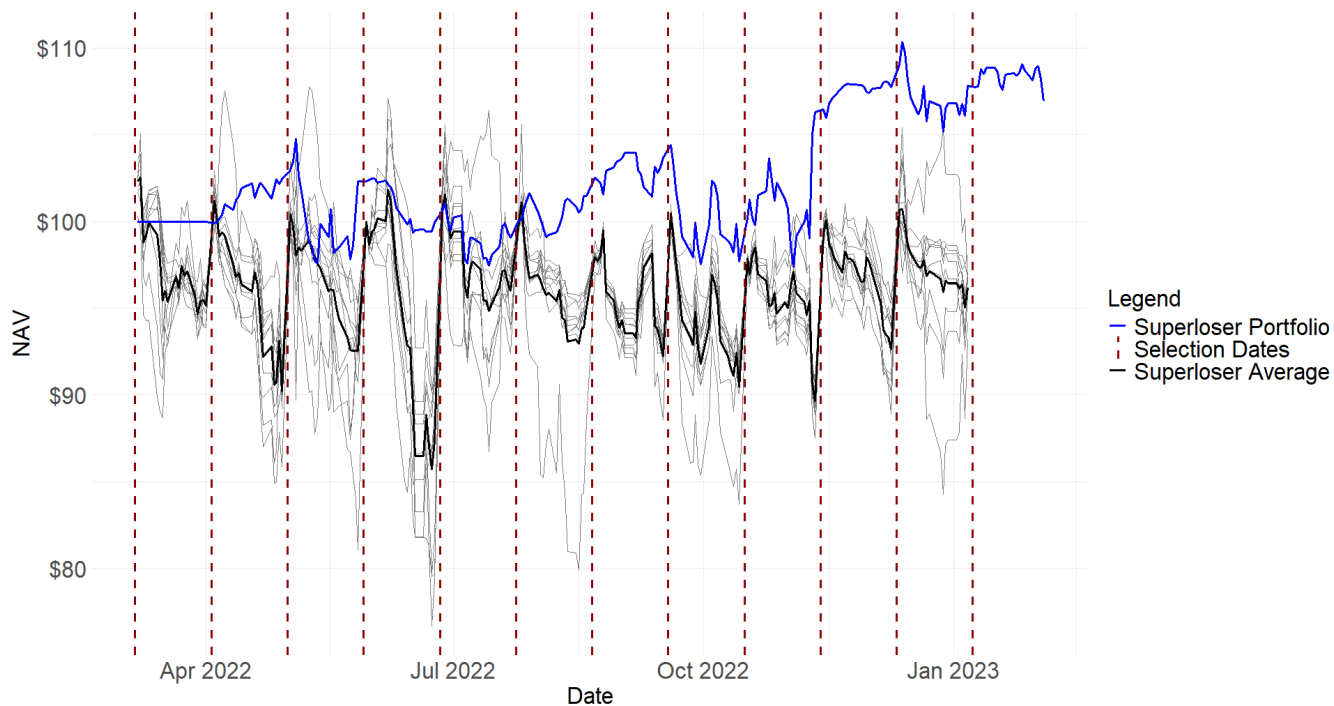

Figure 9: Superlosers strategy: Investing in the bottom 10 losers of the prior month

The unexpected performance of Superlosers vs. Superstars presents a challenge for investors who manage FoFs or invest in active investment managers. Many investors who engage in or believe in active investment are familiar with the concept of momentum investing and may apply such an approach to investing in managers. While momentum has been described as a potentially successful strategy for stocks (Jegadeesh & Titman (1993)), futures contracts (Moskowitz et al. (2012)), broad asset classes via long-only investment in ETFs (Faber, 2007), and almost everything that is liquid and tradeable (Asness et al. (2013)), it does not appear to extend to active investment strategies.

The performance of the Superloser portfolio does not, on its own, suggest that allocation to active strategies is a simple manner of buying the worst performers each month, and even if one desired to apply such a strategy, it is not practical in many cases. While ETFs and mutual funds may be traded in and out of frequently (barring trading restrictions in mutual funds), most hedge funds do not allow frequent trading and often only allow investors to add, subtract, or redeem in entirety on a quarterly basis. Most hedge funds also require a minimum of one month's notice before making a change.

Section 4. Conclusion

M6 competitors were incentivized for extreme performance to win part of the $300,000 prize pool. They held long and short positions on 100 assets selected by the organizers from a clustering solution of

average stock price, coefficient of variation, standard deviation, volume, and total returns. We found that this pool of assets performed better than industry-standard benchmarks such as the S&P-500 and the Nasdaq-100.  Compared to randomly selected portfolios, we found that some competitors had more extreme losses, but only random portfolios that held both long and short positions could exhibit similar extreme performance to competitors while controlling for extreme losses.  We also found that the short competition made it difficult for competitors to demonstrate a positive value-add above industry-standard benchmarks, but most of the significant competitors had negative value-adds. Furthermore, competitors that had returns more similar to benchmark indices were less likely to demonstrate extreme performance in terms of SR (or IR).

The M6 competition highlighted multiple challenges that face investment managers and investors who seek to allocate to active investment managers.  For investment managers, it is quite difficult to beat portfolios comprised of allocations that track simple equal-weight allocations to broad market indices, even over a short period, with a small universe representing only a fraction of the possible assets available.  For end investors, it is very difficult to identify skill amongst investment managers.  Perhaps more importantly, we see direct evidence that picking the managers with the best recent performance, an approach that echoes that of both retail traders as well as the momentum anomaly present in almost every liquid asset class, appears to be a recipe for underperformance.

The authors appreciate the opportunity to review the results of the competition and offer our congratulations and respect to the organizers and participants.

Section 5. References